\documentclass[fleqn,usenatbib,useAMS]{mnras}

\usepackage{graphicx}	
\usepackage{amsmath}	
\usepackage{amssymb}	
\usepackage{multicol}        
\usepackage{bm}		
\usepackage{pdflscape}	

\usepackage{latexsym}
\usepackage{amsmath}
\usepackage{graphicx}
\usepackage{longtable,booktabs}

\title{Classification of Fermi Gamma-Ray Bursts Based on Machine Learning}

\author[Zhu et al.]{Si-Yuan Zhu$^{1}$, Wan-Peng Sun$^{1}$, Da-Ling Ma$^{1}$, Fu-Wen Zhang$^{1,2}$%
	\thanks{Contact e-mail: \href{mailto:fwzhang@pmo.ac.cn}{fwzhang@pmo.ac.cn}}%
	\\
	$^{1}$College of Physics and Electronic Information Engineering, Guilin University of Technology, Guilin 541004, China\\
    $^{2}$Key Laboratory of Low-dimensional Structural Physics and Application, Education Department of Guangxi Zhuang Autonomous Region, \\ Guilin 541004, China}

\date{Last updated xx xx xx; in original form xx xx xx}

\pubyear{2023}

\begin{document}
	\label{firstpage}
	\pagerange{\pageref{firstpage}--\pageref{lastpage}}
	\maketitle
\begin{abstract}
Gamma-ray bursts (GRBs) are typically classified into long and short GRBs based on their durations. However, there is a significant overlapping in the duration distributions of these two categories. In this paper, we apply the unsupervised dimensionality reduction algorithm called t-SNE and UMAP to classify 2061 Fermi GRBs based on four observed quantities: duration, peak energy, fluence, and peak flux. The map results of t-SNE and UMAP show a clear division of these GRBs into two clusters. We mark the two clusters as GRBs-I and GRBs-II, and find that all GRBs associated with supernovae are classified as GRBs-II. It includes the peculiar short GRB 200826A, which was confirmed to originate from the death of a massive star. Furthermore, except for two extreme events GRB 211211A and GRB 230307A, all GRBs associated with kilonovae fall into GRBs-I population. By comparing to the traditional classification of short and long GRBs, the distribution of durations for GRBs-I and GRBs-II do not have a fixed boundary. We find that more than 10\% of GRBs-I have a duration greater than 2 seconds, while approximately 1\% of GRBs-II have a duration shorter than 2 seconds.

\end{abstract}

\begin{keywords}
	gamma-ray burst: general -- stars: statistics
\end{keywords}



\section{Introduction} \label{sec:introduction}

Gamma-ray bursts (GRBs) are the most powerful explosions in the universe \citep{2009ARA&A..47..567G}. Traditionally, GRBs are classified into two categories based on the bimodal distribution of durations: long GRBs (LGRBs) with a duration longer than 2 seconds ($T_{90}>2$ s), and short GRBs (SGRBs) with $T_{90}<2$ s \citep{1993ApJ...413L.101K}. Several lines of observational evidences suggest that some LGRBs are associated with Type Ic supernovae (SNe), such as GRB 980425/SN 1998bw and GRB 030329/SN 2003dh, which is believed to originate from the core-collapse of massive stars \citep{1993ApJ...405..273W,1998Natur.395..670G,2003ApJ...591L..17S,2006ARA&A..44..507W,2012grb..book..169H}.
Recently, the discovery of GRB 170817A that are associated with gravitational wave (GW) GW170817 and kilonova (KN) AT 2017gfo, has confirmed that a part of SGRBs originate from mergers of binary compact star \citep{2017PhRvL.119p1101A,2017ApJ...848L..14G,2017ApJ...848L..15S,2017ApJ...851L..18W}.

However, the traditional classification schemes have been challenged by recent observations, the dichotomy based on phenomenon does not necessarily correspond to the two distinct physical origins of GRBs. Some short-duration GRBs (such as GRB 090426, GRB 100816A, and GRB 200826A) are thought to possibly originate from massive collapsars \citep{2009ApJ...703.1696Z,2011ApJ...739...47F,2011A&A...531L...6N,2011MNRAS.410...27X,2012ApJ...750...88Z,
	2021NatAs...5..917A,2021NatAs...5..911Z,2022ApJ...932....1R}, while some long-duration GRBs (such as GRB 060505, GRB 060614, 211227A, and GRB 211211A) are believed to possibly originate from compact binary mergers \citep{2006Natur.444.1050D,2006Natur.444.1053G,2006Natur.444.1044G,2006Natur.444.1047F,2007ApJ...655L..25Z, 2022ApJ...931L..23L,2022Natur.612..223R,2022Natur.612..228T,2022arXiv220502186X,2022Natur.612..232Y,2022ApJ...936L..10Z}.
Moreover, a significant overlapping is presented in the $T_{90}$ distribution of SGRBs and LGRBs, and the measurement of $T_{90}$ strongly depends on instrument and energy band \citep{2012ApJ...750...88Z,2013ApJ...763...15Q}. It means that $T_{90}$ is not necessarily a reliable indicator of the physical nature of a GRB.

Besides $T_{90}$, other observational features have been also used to identify GRBs. For example, spectral lag that is defined as the time delay of high-energy photons with respect to low-energy photons, is an important quantity for the classification of GRBs \citep{1986ApJ...301..213N,1995A&A...300..746C,2006MNRAS.367.1751Y,2015MNRAS.446.1129B}.
In general, LGRBs have significant spectral lags, while SGRBs have negligible or zero spectral lags \citep{2006MNRAS.367.1751Y,2015MNRAS.446.1129B,2017ApJ...844..126S}. Moreover, spectral hardness can usually be represented by hardness ratio (HR), spectral index ($\alpha$), and peak energy ($E_{\rm p}$) in the $\nu f_\nu$ spectrum of the prompt emission.
Generally, the spectrum of SGRBs are harder than that of LGRBs, and it can be demonstrated through the HR--$T_{90}$ plane and $E_{\rm p}$--$T_{90}$ plane \citep{2012ApJ...750...88Z}.
The rest-frame peak energy--isotropic energy ($E_{\rm p,z}$--$E_{\rm iso}$) relation is also a widely discussed GRB classifier, even though the dispersion of the correlation is large, and outliers do exist \citep{2012ApJ...750...88Z,2013MNRAS.430..163Q,2020ApJ...903L..32Z,2021NatAs...5..877A,2021NatAs...5..911Z,2022Natur.612..232Y,2022ApJ...936L..10Z,2023ApJ...950...30Z}.
\cite{2010ApJ...725.1965L} and \cite{2020MNRAS.492.1919M} proposed two similar parameters to classify GRBs, $\varepsilon = (E_{\rm iso}/10^{52})/(E_{\rm p,z}/100)^{5/3}$ and energy--hardness ($EH = (E_{\rm p,z}/100)/(E_{\rm iso}/10^{51})^{0.4}$), respectively.
Subsequently, \cite{2020MNRAS.492.1919M} found a boundary on the $EH$--$T_{90}$ plane, and proposed a new parameter, energy--hardness--duration ($EHD = EH/T_{\rm 90,z}^{0.5}$), to classify GRBs.

According to the generally accepted two-types progenitors of GRBs, the properties of their host galaxies are statistically different \citep{2009ApJ...703.1696Z,2010ApJ...708....9F,2016ApJS..227....7L}. GRBs that are originated from mergers of compact stars, are usually located in the faint regions of the dwarf and elliptical galaxies, and they have a large offset from the galaxy center with a small local star formation rate \citep{2005Natur.438..988B,2005Natur.437..851G,2010ApJ...708....9F}.
However, GRBs originating from collapsars are usually located at the bright regions in the irregular and dwarf galaxies which have small offset from the center and a large local star formation rate.
According to the multi-band observations of GRBs including their host galaxies, \cite{2009ApJ...703.1696Z} classified GRBs as Type I and Type II which are corresponding to the progenitors of massive star collapsar and compact star merger, respectively.

In addition to these two main classes explored above in detail, a number of sub-classes have also been proposed, such as Ultra-long GRBs, X-ray flashes/X-ray rich GRBs, SGRBs with extended emission, and so on, even though their progenitors have not yet been identified. Furthermore, some authors suggested that there are three or five types of GRBs \citep{1998ApJ...508..757H,1998ApJ...508..314M,2017MNRAS.469.3374C,2018MNRAS.475.1708A}. \cite{2023MNRAS.525.5204B} applied the Gaussian Mixture Model (GMM) to explore whether there is more GRB sub-classes based on a broader set of parameters, including prompt and plateau
emission ones. They found the microtrends of sub-classes. For more details please refer to \cite{2023MNRAS.525.5204B} and the references therein.

Recently, a large number of mathematical and/or machine learning methods have also been applied to the GRB classification, such as K-means cluster, t-mixtures-model-based cluster (tMMBC), principal component analysis (PCA), naive bayes, fuzzy cluster, and XGBoost \citep{2007ApJ...667.1017C,2016MNRAS.462.3243Z,2017MNRAS.469.3374C,2018MNRAS.475.1708A,2018MNRAS.481.3196C,
	2019MNRAS.486.4823T,2020ApJ...897..154L,2021A&C....3400441M,2023ApJ...959...44L,2023MNRAS.525.5204B}.
In addition, the unsupervised dimensionality reduction algorithm, t-distributed stochastic neighbour embedding (t-SNE) \citep{2008JMLR.9.2579M,2014JMLR.15.3221M} and Uniform Manifold Approximation and Projection (UMAP) \citep{2018arXiv180203426M} have been widely used to do the classification of various types of objects. \cite{2020ApJ...905...97Z} used t-SNE method to classify type-2 active galactic nucleus (AGN) and HII galaxies based on four physical quantities.
\cite{2022MNRAS.509.1227C} and \cite{2023MNRAS.519.1823Z} used UMAP to classify fast radio bursts (FRBs) based on ten physical quantities. Both t-SNE and UMAP have been also used to classify GRBs based on Swift's lightcurve \citep{2020ApJ...896L..20J,2023ApJ...949L..22D,2023ApJ...945...67S}.

Motivated by the recent observational advances of GRBs, we reanalyze the classification of Fermi GRBs using the t-SNE and UMAP methods. The four key physical parameters (duration, peak energy, fluence and peak flux) are adopted. The structure of our paper is organized as follows.
In Section \ref{sec:method}, we describe the t-SNE and UMAP methods, as well as the sample selection.
A clear classification based on the t-SNE and UMAP map and the statistical analysis for this classification is shown in Section \ref{sec:classification}.
The distribution of some special GRBs on t-SNE and UMAP maps and the discussions of the physical meaning of this new classification are shown in Section \ref{sec:discussions}.
The conclusions are shown in Section \ref{sec:conclusions}.
The symbolic notation $Q_{\rm n} = Q/10^{\rm n}$ is adopted.

\section{Method and Data} \label{sec:method}
\subsection{t-Distributed Stochastic Neighbour Embedding} \label{subsec:t-sne}
t-SNE is an unsupervised machine learning algorithm that can nonlinearly reduce high-dimensional data to two-dimensional or three-dimensional for visualization \citep{2008JMLR.9.2579M,2014JMLR.15.3221M}.
The basic principle of t-SNE is to establish a mapping relationship between high-dimensional space and low-dimensional space, so that similar data points in high-dimensional space are embedded in similar positions in low-dimensional space, while dissimilar data points in high-dimensional space are embedded in far positions in low-dimensional space.
The calculation of the axes in the embedded low-dimensional space depends on the similarity between sample points in the high-dimensional space and the distance between sample points in the low-dimensional space.
Furthermore, this process depends upon random initialization, and running t-SNE on an identical dataset can produce a variety of maps with similar topologies \citep{2020ApJ...891..136S,2020ApJ...905...97Z}.
Thus, the axes resulting from t-SNE dimensionality reduction do not have proper labels or physical meaning, and only represent the structure and distribution of the data in the high-dimensional space, called the t-SNE dimension x and the t-SNE dimension y.

The most important hyperparameters applied in the t-SNE technique (the hyperparameters listed in sklearn.manifold.TSNE in python) is the $perplexity$, which determines the sizes of the neighbourhoods based on the density of the data in the respective regions and can be approximately interpreted as the typical number of neighbours which should be considered similar when computing distances.
A higher $perplexity$ value will consider a larger number of neighbors, emphasizing the global structure of the data.
Conversely, a lower $perplexity$ value will better reflect the particular structure of the data, and choosing the appropriate $perplexity$ value is crucial for representing both local and global aspects of the data. It should be noted that t-SNE requires that each object in the dataset have a uniform data dimension without any missing values.

\subsection{Uniform Manifold Approximation and Projection} \label{subsec:umap}
UMAP is a non-linear dimension reduction algorithm based on manifold learning techniques and topological data analysis, and is also an unsupervised machine learning algorithm. It is similar to t-SNE, but UMAP can preserve the integrity of global data more completely than t-SNE in theory and implementation (Due to the complexity of t-SNE calculations, the dataset usually uses PCA to reduce the dimensionality first), and has a better dimensionality reduction effect \citep{2008JMLR.9.2579M,2014JMLR.15.3221M,2018arXiv180203426M}.
UMAP also embeds high-dimensional data into low-dimensional space while retaining the global similar structure. Similarly, the axes of this low-dimensional space do not have proper labels or physical meaning.

We use a python module umap-learn\footnote{https://pypi.org/project/umap-learn/} introduced by \citet{2018arXiv180203426M} to obtain the UMAP analysis results. The most important hyperparameters of UMAP are $n\_neighbors$ and $min\_dist$.
$n\_neighbors$ determine the number of neighboring points used in local approximations of manifold structure, and $min\_dist$ controls how tightly the embedding is allowed to compress points together. Note that when UMAP is applied to the dataset, the dimensions of each object must also be the same.

\subsection{Data} \label{subsec:data}

The Fermi GRBs are taken from the Fermi Catalog\footnote{https://heasarc.gsfc.nasa.gov/W3Browse/fermi/fermigbrst.html.} until the end of April 2021. In total, there are 3029 events.
We mainly consider GRBs with the well-measured four-dimensional data including the duration ($T_{90}$), peak energy ($E_{\rm p}$), peak flux ($F_{\rm p}$) and fluence ($S_{\gamma}$).
In order to eliminate the effect of the time resolution on the peak flux, we uniformly select the $F_{\rm p}$ with the timescale of 64 $\rm ms$ for all GRBs.
The four spectral models including the Band model \citep{1993ApJ...413..281B}, the power law (PL) model, the cutoff power law (CPL) model, and the smoothly broken power law (SBPL) model are used to fit the spectra of GRBs.
The values of $E_{\rm p}$ and $S_{\gamma}$ are mainly taken from the best spectral fitting model.
If the best fitting model of one GRB is PL model, we take the CPL model to obtain the $E_{\rm p}$ value.
To ensure the accuracy of $E_{\rm p}$ in our sample, we exclude the GRBs in which the error of $E_{\rm p}$ is larger than 40\%.
A total of 2057 GRBs with well four-dimensional data are selected.

There are several unique GRBs, including GRB 190114C, GRB 200826A, GRB 211211A, and GRB 230307A.
Although they have no time-integrated spectrum data in the Fermi Catalog, we still include these GRBs in our sample and take the values of $E_{\rm p}$ and $S_{\gamma}$ from the GRB Coordinates Network (GCN).
Finally, we compile a Fermi sample containing 2061 GRBs.

\section{Classification} \label{sec:classification}

\begin{table*}
	\caption{The prompt emission parameters and classification results of Fermi GRBs based on t-SNE and UMAP methods}
	\label{t-fm}
	\tiny
	\begin{tabular}{lcccccc}
		\hline
		$GRB$ & $T_{90}$ & $E_{\rm p}$  & $S_{\gamma,-6}$ & $F_{\rm p}$ & t-SNE$^{a}$ & UMAP$^{a}$
		\\
		& (s) & (keV)  & (erg cm$^{-2}$) & (ph cm$^{-2}$ s$^{-1}$) & &
		\\
		\hline
		GRB080714086 & 5.38 & 364.33 & 1.03 & 3.82 & II & II \\
		GRB080714745 & 59.65 & 214.89 & 4.82 & 8.89 & II & II \\
		GRB080715950 & 7.87 & 270.4 & 5.32 & 19.42 & II & II \\
		GRB080717543 & 36.61 & 627.11 & 4.47 & 6.24 & II & II \\
		GRB080723557 & 58.37 & 213.97 & 83.5 & 40.97 & II & II \\
		GRB080723913 & 0.19 & 320.84 & 0.23 & 5.26 & I & I \\
		GRB080723985 & 42.82 & 429.8 & 35.69 & 13.45 & II & II \\
		GRB080724401 & 379.4 & 102.67 & 11.68 & 22.73 & II & II \\
		GRB080725435 & 25.92 & 372.68 & 9.37 & 5.38 & II & II \\
		GRB080725541 & 0.96 & 1482.59 & 0.81 & 6.27 & I & I \\
		...\\
		\hline
		\multicolumn{5}{l}{\bf Notes.}\\
		\multicolumn{5}{l}{(a) {I and II represent GRBs-I and GRBs-II, respectively.}}\\
		\multicolumn{5}{l}{(This table represents a small portion of our Fermi GRB sample, the full table is available in machine-readable form.)}\\
	\end{tabular}
\end{table*}

\begin{figure*}
	\centering
	\includegraphics[angle=0,scale=0.52]{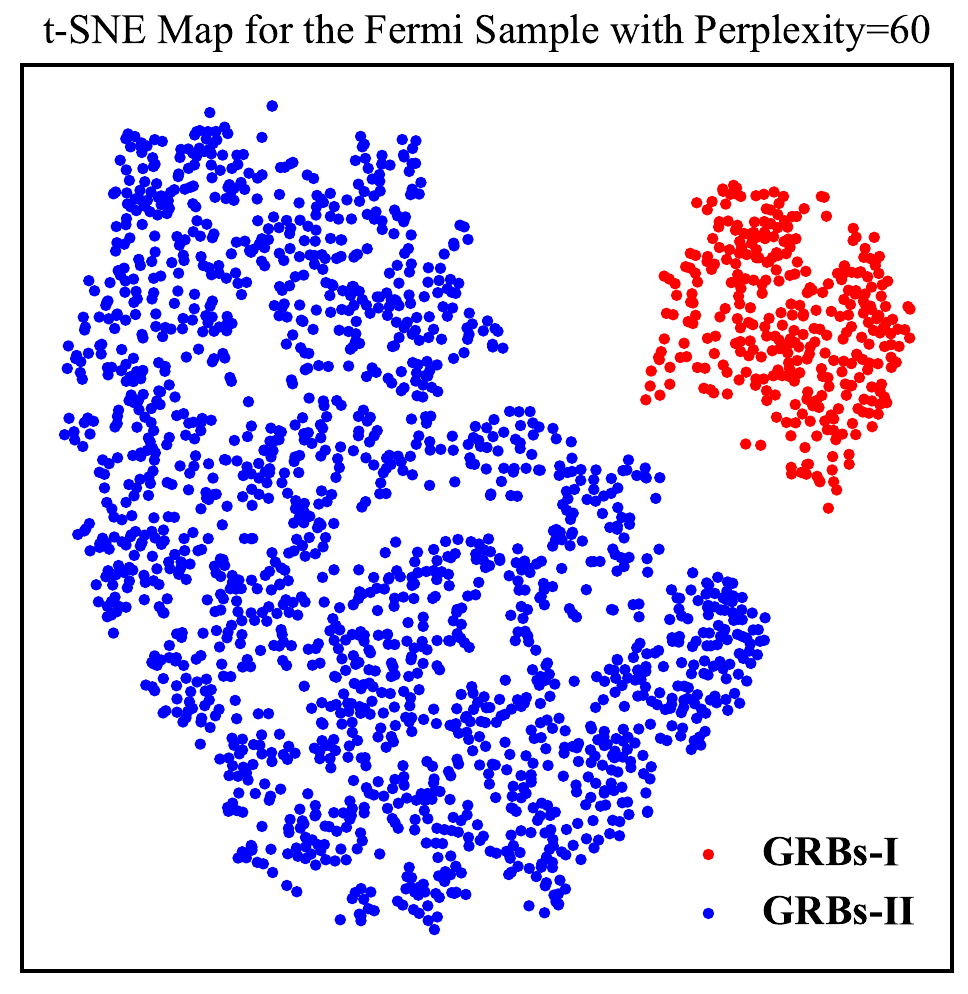}
	\includegraphics[angle=0,scale=0.52]{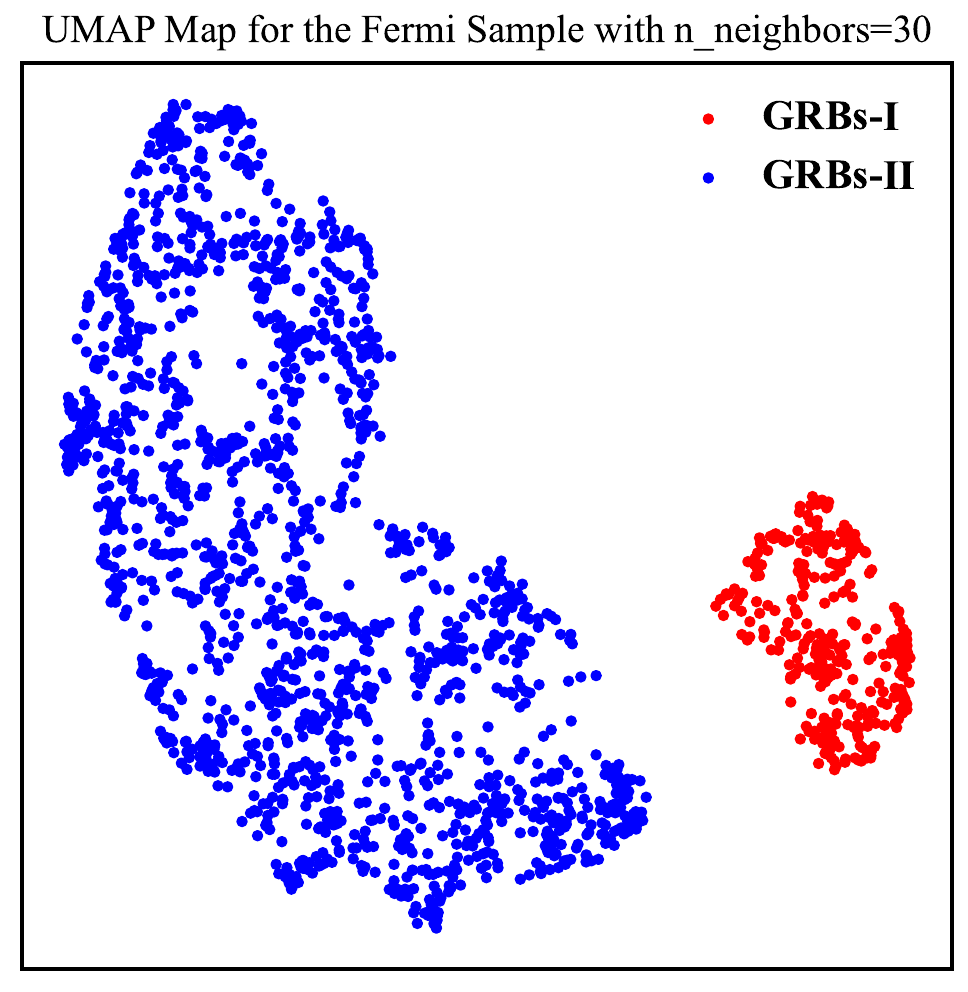}
	\caption{The t-SNE and UMAP 2D projection of the 2061 GRBs from the Fermi Catalog based on $T_{90}$, $E_{\rm p}$, $S_{\gamma}$, and $F_{\rm p}$. There are clear two clusters: one cluster with dots in red (GRBs-\uppercase\expandafter{\romannumeral1)} and the other cluster with dots in blue (GRBs-\uppercase\expandafter{\romannumeral2)}. The axes resulting from the t-SNE and UMAP have no clear physical interpretation or units, and only the structures are meaningful.}
	\label{f-classification}
\end{figure*}

\begin{figure*}
	\centering
	\includegraphics[angle=0,scale=0.52]{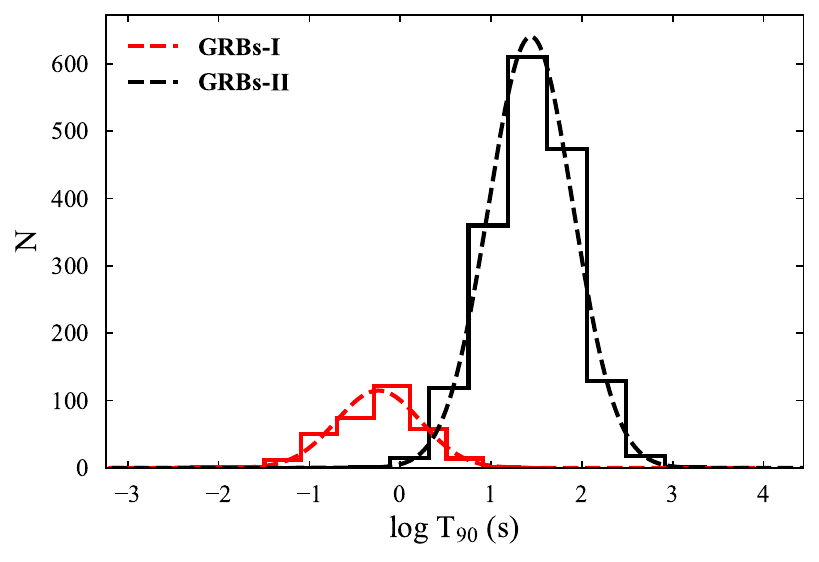}
	\includegraphics[angle=0,scale=0.52]{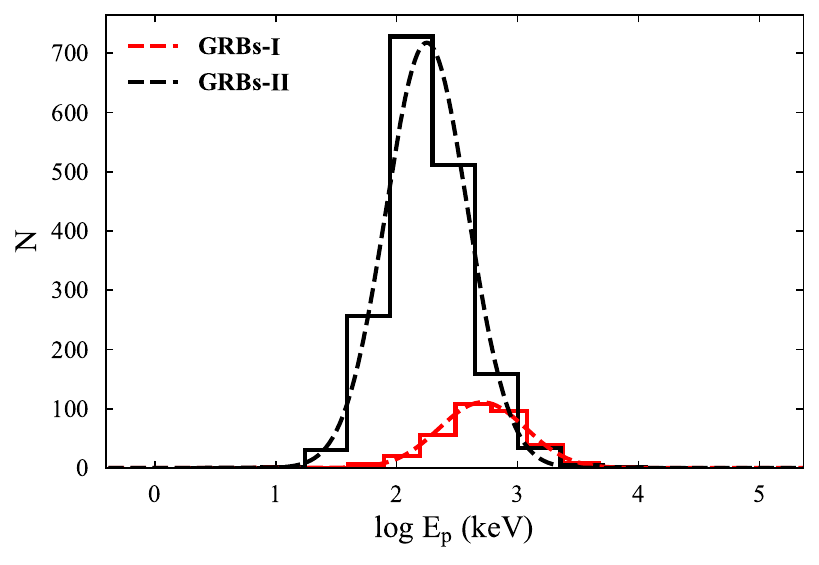}
	\includegraphics[angle=0,scale=0.52]{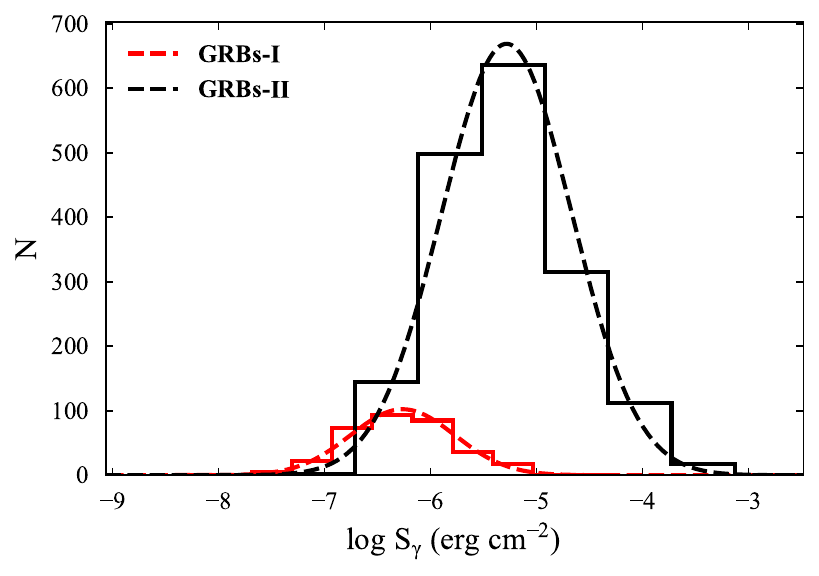}
	\includegraphics[angle=0,scale=0.52]{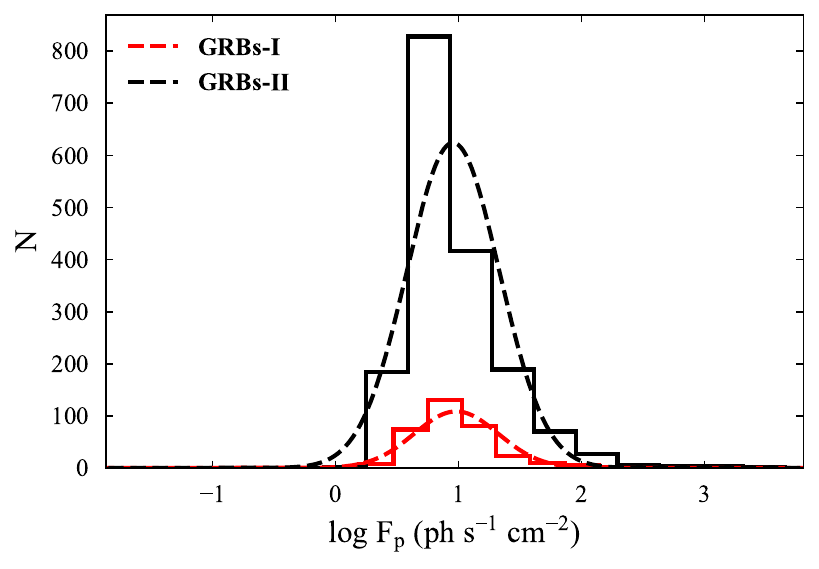}
	\caption{Distributions of $T_{90}$, $E_{\rm p}$, $S_{\gamma}$, and $F_{\rm p}$ in the observer frame based on the UMAP method. The dotted lines represent Gaussian fitting curves.}
	\label{f-distributions}
\end{figure*}

We use t-SNE ($perplexity=60$) and UMAP ($n\_neighbors=30$ and $min\_dist=0.01$) methods to map the Fermi sample based on $T_{90}$, $E_{\rm p}$, $S_{\gamma}$, and $F_{\rm p}$, respectively, as shown in Figure \ref{f-classification}. Interestingly, the GRBs of the Fermi sample are clearly divided into two clusters, between which one is larger and the other is smaller, and t-SNE and UMAP maps present similar structures.
To comply with traditional classification methods, we call the small cluster as GRBs-I and the larger cluster as GRBs-II. The detailed classification result and the prompt emission parameters of Fermi GRBs based on t-SNE and UMAP methods are listed in Table \ref{t-fm}.
We also investigate the difference between results of t-SNE and UMAP, and find that the types of 7 GRB are different from the results of t-SNE and UMAP, where 6 GRBs (GRB 081130212, GRB 090320045, GRB 101002279, GRB 120504945, GRB 131128629, GRB 140912664) are classified as GRBs-I on the t-SNE map, while they are classified as GRBs-II on the UMAP map; and one GRB (GRB 150819440) is classified as GRBs-II on the t-SNE map, while which is classified as GRBs-I on the UMAP map.
For UMAP (t-SNE) result, there are 334 (339) GRBs-I, accounting for 16.2\% (16.4\%) of the Fermi sample, and 1727 (1722) GRBs-II, accounting for 83.8\% (83.6\%) of the Fermi sample.
Since there is only a few difference between t-SNE and UMAP, we mainly focus on the UMAP result in the following text.
The $T_{90}$, $E_{\rm p}$, $S_{\gamma}$, and $F_{\rm p}$ distributions based on UMAP method are shown in Figure \ref{f-distributions}.
For GRBs-I (GRBs-II), the median values and dispersions are $T_{90} \sim 0.58~ (27.38)$ s, $\sigma \sim 0.46~ (0.47)$, $E_{\rm p} \sim 512~ (177)$ keV, $\sigma \sim 0.36~ (0.34)$, $S_{\gamma} \sim 0.55~ (5.24) \times 10^{-6}$ erg $\rm cm^{-2}$, $\sigma \sim 0.49~ (0.62)$, $F_{\rm p} \sim 9.60~ (9.04)$ ph $\rm s^{-1}$ $\rm cm^{-2}$, $\sigma \sim 0.34~ (0.38)$.

We find that GRBs-I have relatively short $T_{90}$.
The median values of the $T_{90}$ distribution of the two clusters are significantly separated, and the $T_{90}$ of the two clusters show bimodal distribution as a whole.
For the distributions of $E_{\rm p}$, $S_{\gamma}$, and $F_{\rm p}$, there are no obvious distinctions between GRBs-I and GRBs-II, but the median values of $E_{\rm p}$ and $F_{\rm p}$ of GRBs-I are larger than GRBs-II, and the median value of $S_{\gamma}$ of GRBs-I is smaller than that of GRBs-II.
The distribution characteristics $T_{90}$, $E_{\rm p}$, $S_{\gamma}$, and $F_{\rm p}$ of GRBs-I and GRBs-II are similar to those of the previous SGRBs and LGRBs.

\begin{figure*}
	\centering
	\includegraphics[angle=0,scale=0.31]{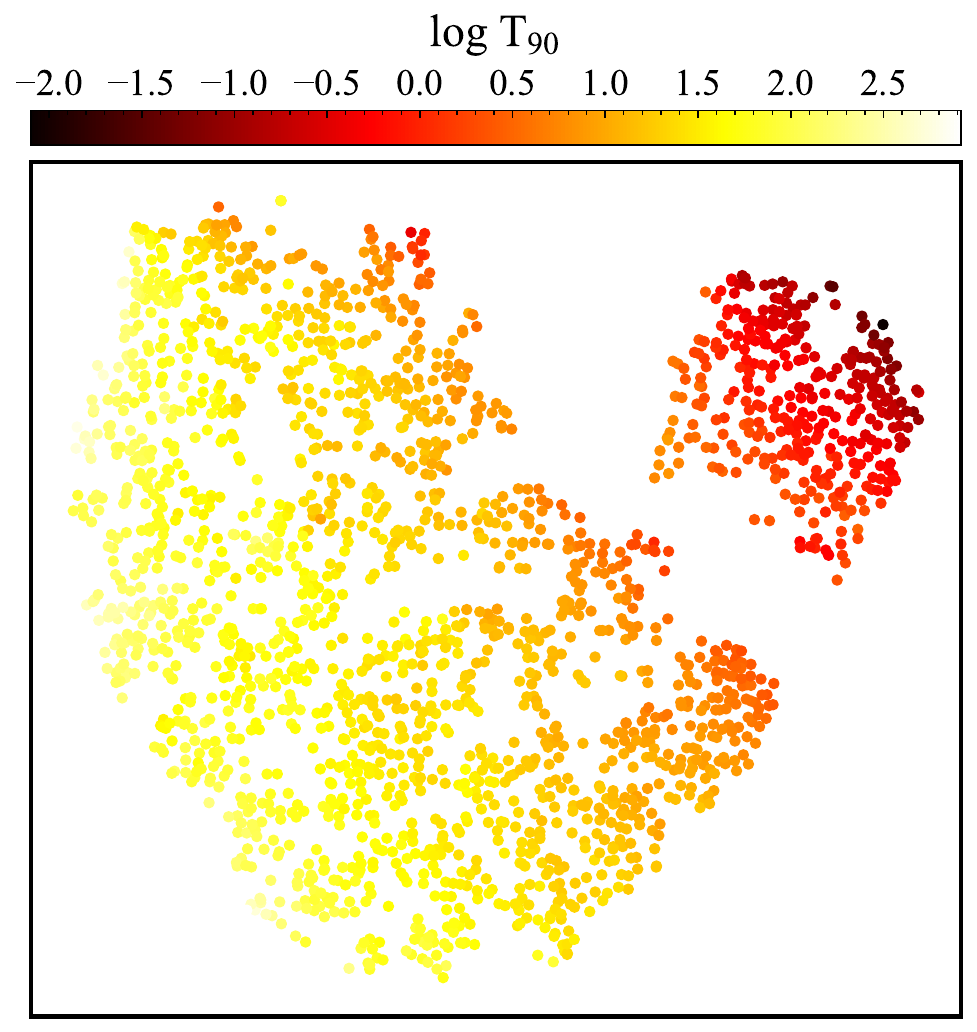}
	\includegraphics[angle=0,scale=0.31]{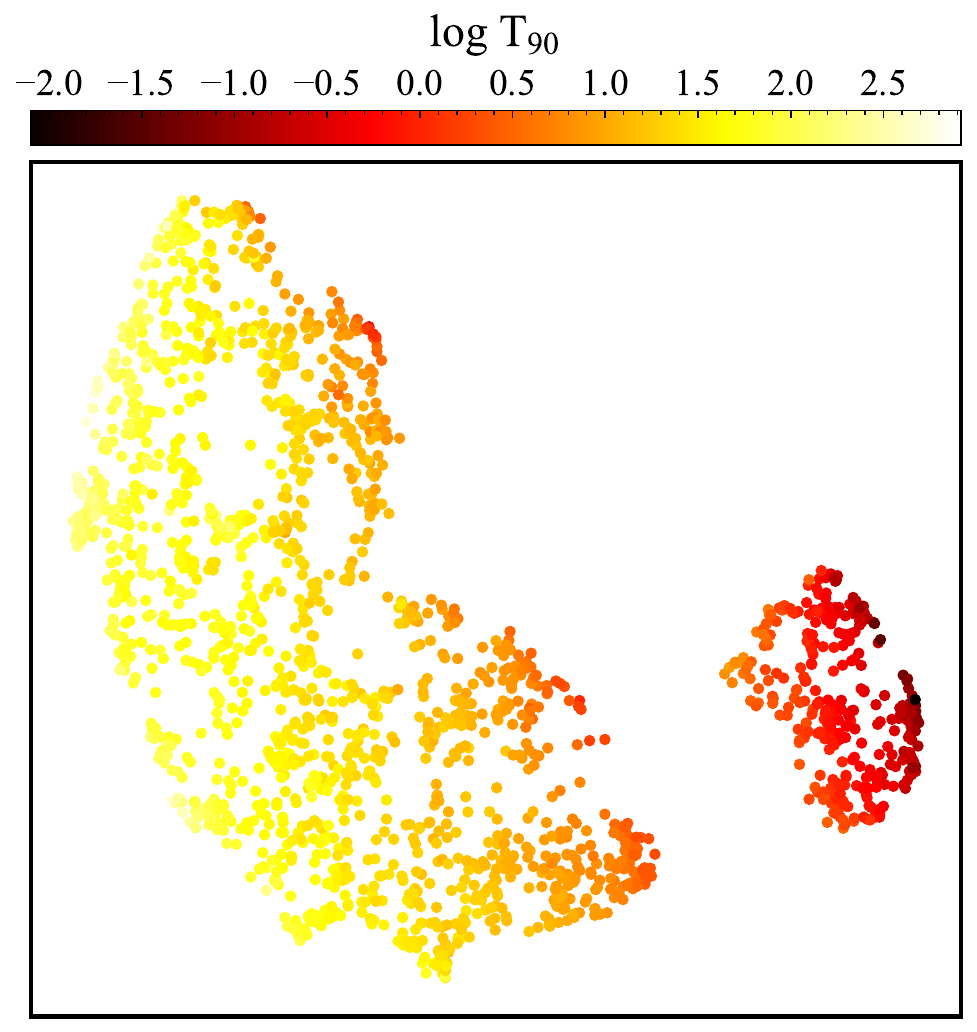}
	\\
	\includegraphics[angle=0,scale=0.31]{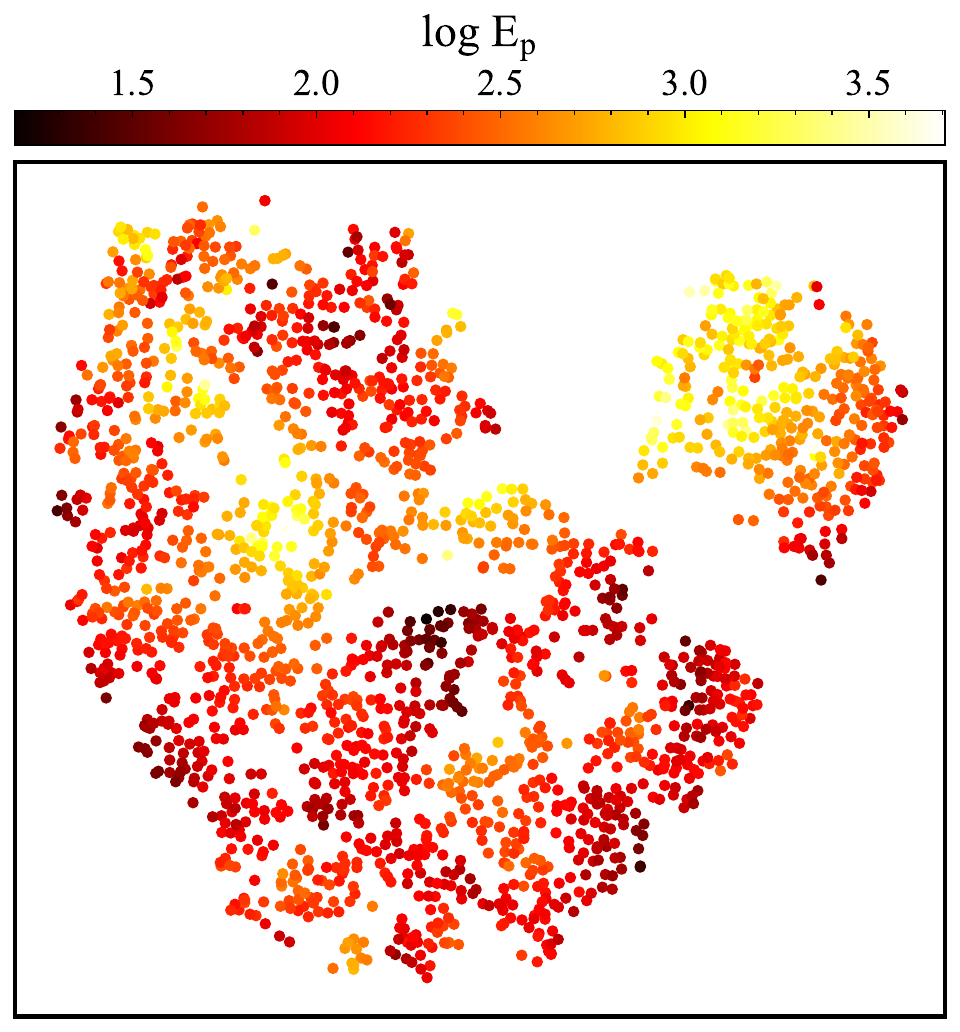}
	\includegraphics[angle=0,scale=0.31]{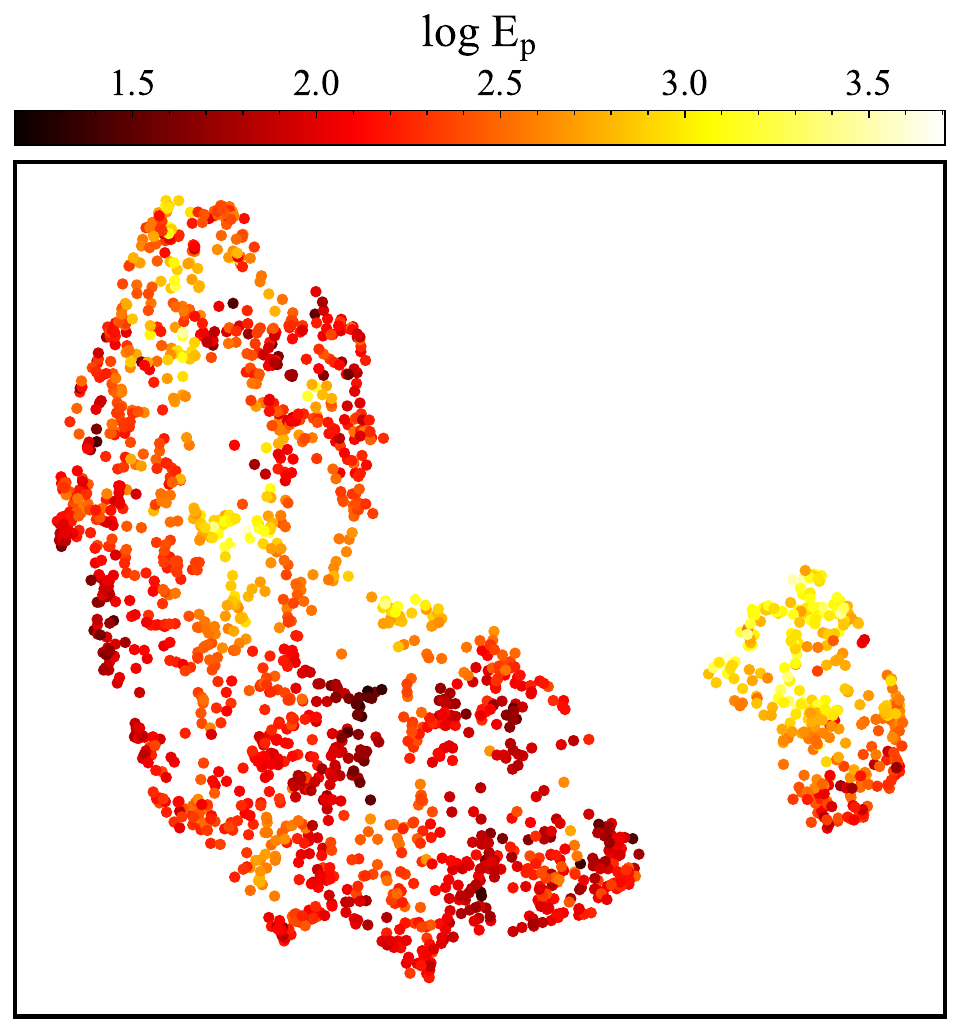}
	\\
	\includegraphics[angle=0,scale=0.31]{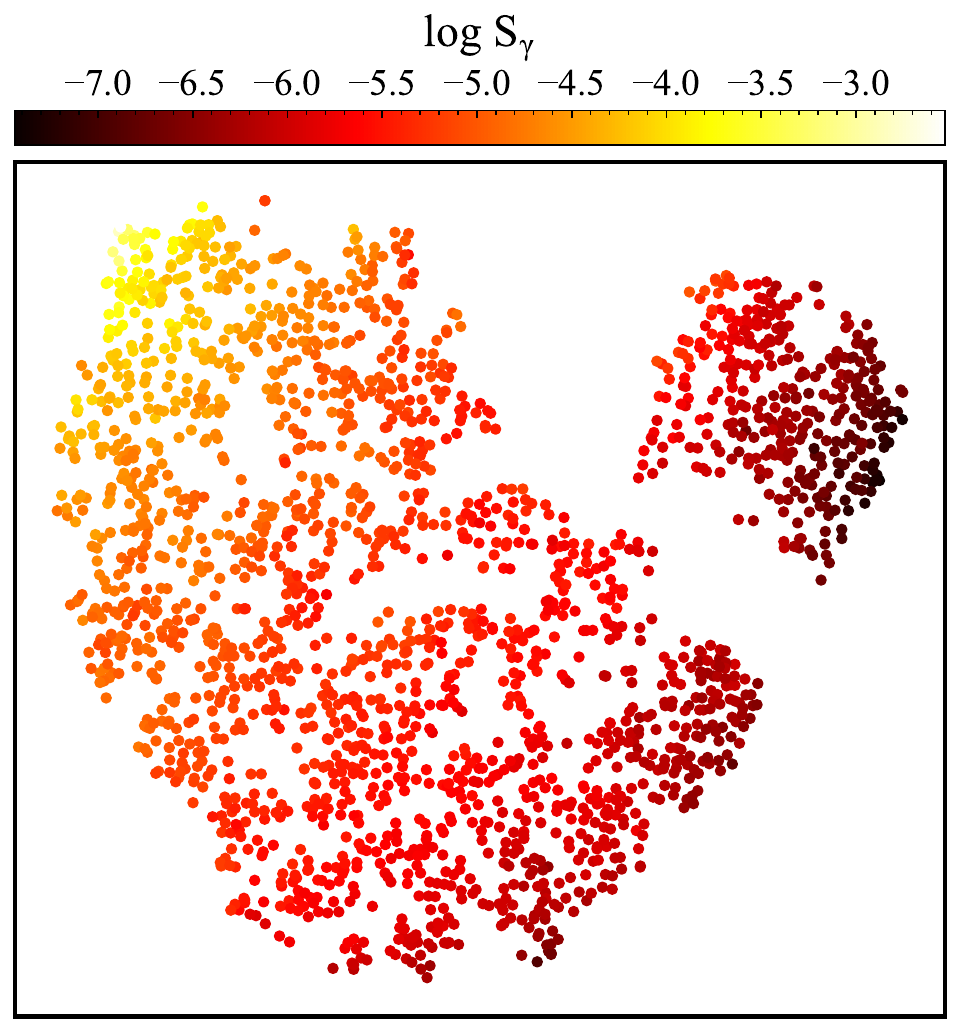}
	\includegraphics[angle=0,scale=0.31]{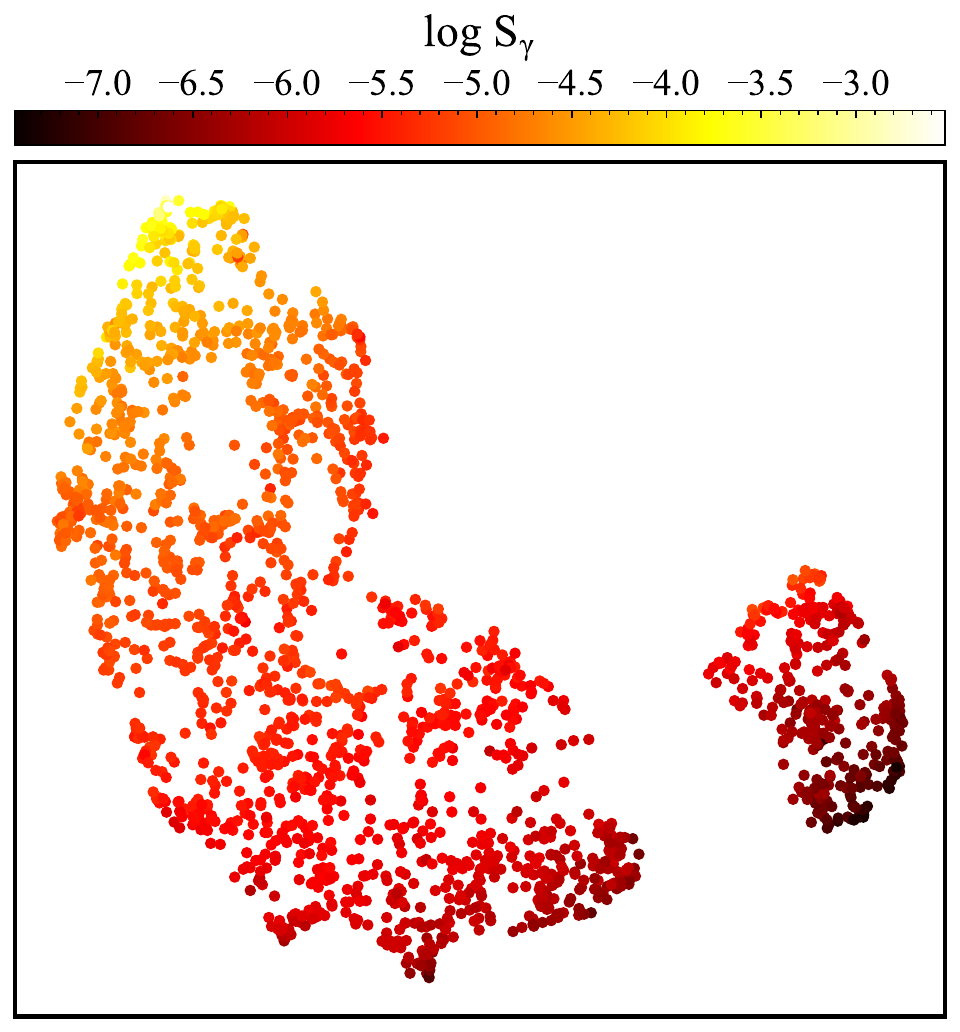}
	\\
	\includegraphics[angle=0,scale=0.31]{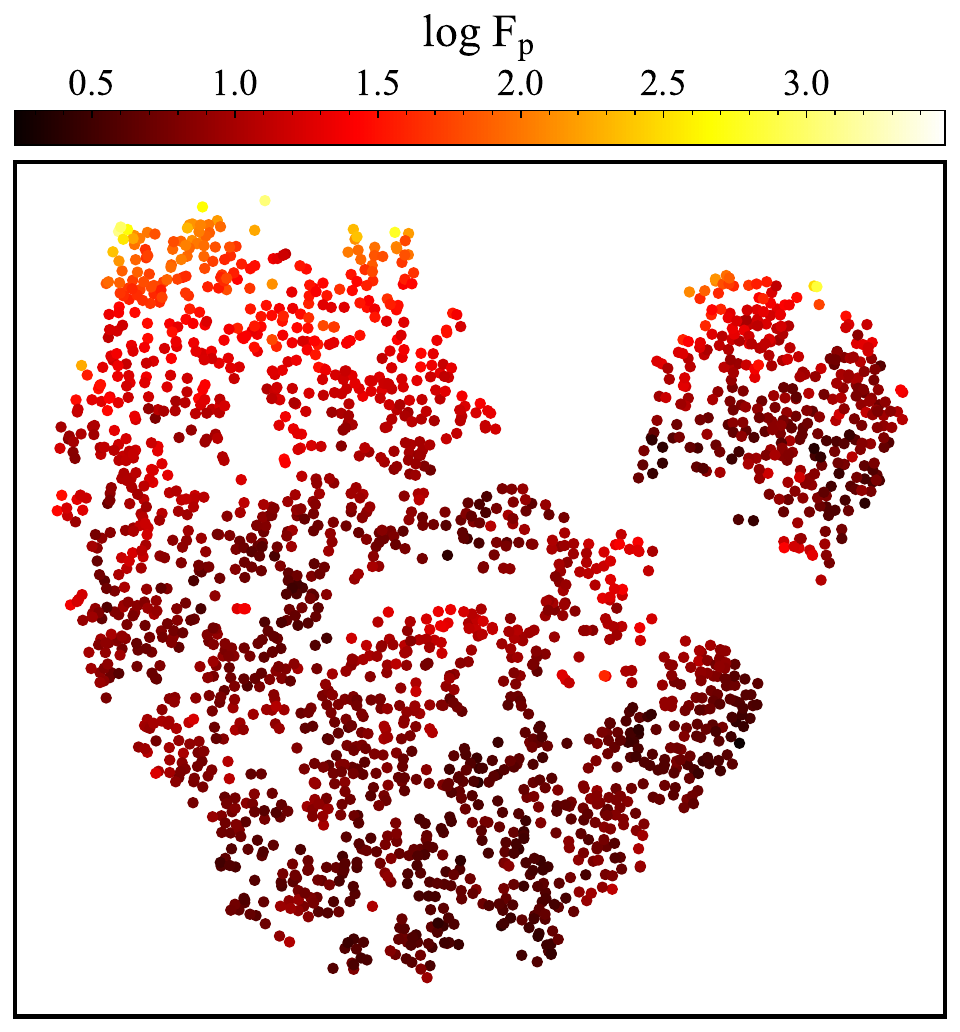}
	\includegraphics[angle=0,scale=0.31]{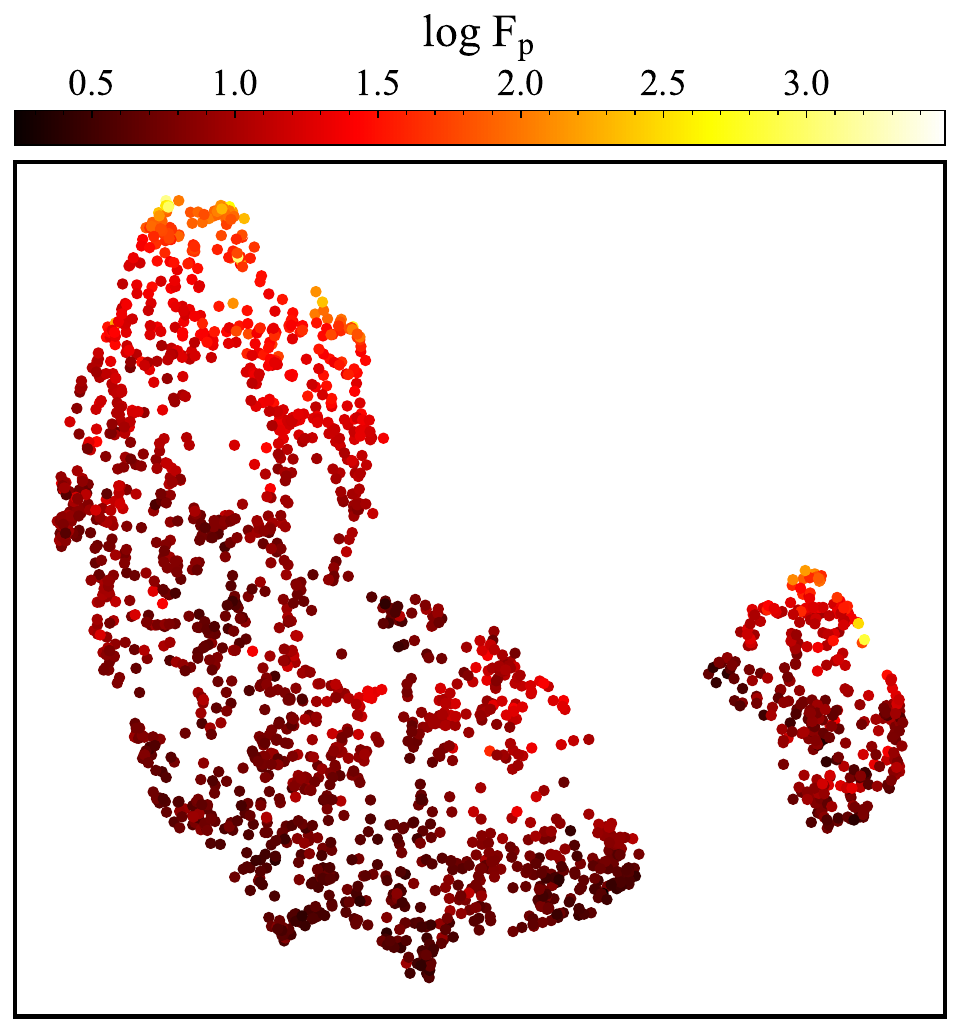}
	\caption{t-SNE (on the left) and UMAP (on the right) maps of Fermi GRBs, colored based on log $T_{90}$, log $E_{\rm p}$, log $S_{\gamma}$, and log $F_{\rm p}$, respectively.}
	\label{f-color}
\end{figure*}

We color each GRB with $T_{90}$, $E_{\rm p}$, $S_{\gamma}$, and $F_{\rm p}$ in the two clusters, respectively.
As shown in Figure \ref{f-color}, t-SNE and UMAP maps show similar trends of change.
The distributions of $T_{90}$, $S_{\gamma}$ and $F_{\rm p}$ on the t-SNE and UMAP map will rise gradually with a certain direction, and the color will obviously change from dark to light (from dark red to light yellow).
t-SNE and UMAP have roughly approximated the GRBs into two clusters according to $T_{90}$.
However, there is no absolute boundary for $T_{90}$ alone between the two clusters.
Specifically, $	T_{90}$ of some GRBs-I is longer than that of some GRBs-II.
For GRBs-I, $T_{90}$ can reach a maximum of 8 s, and for GRBs-II, $T_{90}$ can reach a minimum of 0.4 s, which is different from the traditional classification methods (which take $T_{90}=2$ s as the boundary).
We also find that there are some sub-structures in GRBs-I and GRBs-II, but this is not the focus of this work, so we will not do detailed studies in this paper.
Note that, changing the $perplexity$ or $n\_neighbors$ will obtain different topological structures to study more refined sub-structures, but only suitable data can obtain this clearly separated structure, and if the data is not appropriate, no matter how the perplexity is changed, it will not achieve obvious separate structure, which is also not the focus of this work.

\section{Discussions} \label{sec:discussions}
\subsection{GRBs Associated with Other Electromagnetical Counterparts} \label{subsec:origin}

The t-SNE and UMAP maps indeed show that the bursts with similar properties tightly cluster together, and the two clusters may strongly indicate intrinsically different physical properties and/or different origins.
Although the classification of individual GRBs as collapsar or merger based on simple criteria such as duration, light curve and so on, is uncertain, particularly for some confusing bursts, many bursts can be unambiguously classified as collapsar or merger on the basis of other observations such as supernova, kilonova and gravitation wave. It is therefor necessary to confirm that whether the classification proposed here matches the previous results.

It is known that if one GRB is associated with a supernova, it is unambiguously a collapsar origin. 14 GRBs in the Fermi Catalog have clearly been observed in association with supernova.
Among these bursts, 9 GRBs (GRB 091127, GRB 101219B, GRB 130215A, GRB 130427A, GRB 130702A, GRB 140606B, GRB 171010A, GRB 190114C, and GRB 190829A) are well-characterized by spectroscopy (SNsp) \citep{2011ApJ...743..204B,2011ApJ...735L..24S,2013ApJ...776...98X,2014A&A...568A..19C,2014A&A...569A.108K,2015MNRAS.452.1535C,2019MNRAS.490.5366M,2019GCN.23983....1M,2019GCN.25677....1D}, and 5 GRBs (GRB 090618, GRB 111228A, GRB 120729A, GRB 141004A and GRB 200826A) are characterized by photometry (SNph) \citep{2020MNRAS.492.1919M}. Due to the lack of available data for GRB 120729A and GRB 190829A, these two bursts are not included in the analysis.
The remaining 12 bursts associated with SNe are all classified as GRBs-II by the t-SNE and UMAP maps (Figure \ref{f-tsne1}, the key physical parameters are listed in Table \ref{t-special}), especially for GRB 200826A, a peculiar short GRB arisen from a collapsar \citep{2021NatAs...5..917A,2021NatAs...5..911Z,2022ApJ...932....1R}.

In our sample, three GRBs, GRB 150101B, GRB 160821B and GRB 170817A are associated with KNe \citep{2017ApJ...851L..18W,2018NatCo...9.4089T,2019ApJ...883...48L,2019MNRAS.489.2104T}, and all belong to GRBs-I (Figure \ref{f-tsne1}).
GRB 170817A associated with GW170817, has been confirmed that it originates from the merger of binary neutron stars \citep{2017PhRvL.119p1101A,2017ApJ...848L..14G,2017ApJ...851L..18W}.
Recently, it is reported that long duration GRB 211211A is also accompanied by a KN, which might
originate from a compact star merger \citep{2022Natur.612..223R,2022arXiv220502186X,2022Natur.612..232Y}.
This burst was located at the edge of the GRBs-II on the t-SNE and UMAP maps.
However, \cite{2023ApJ...947...55B} suggested that collapsars can also explain the origin of GRB 211211A. Thus the physical origin of GRB 211211A needs to be further explored and confirmed by more observations.
Similarly, some authors reported that GRB 230307A may also arise from a merger, which is here located on the edge of the GRBs-II, and close to GRB 211211A on both t-SNE and UMAP maps. \citep{2024Natur.626..737L,2024Natur.626..742Y}. The main parameters of these bursts are also listed in Table \ref{t-special}.

\begin{table*}
	\caption{The prompt emission parameters of GRBs associated with SN or GW/KN observations in the sample}
	\label{t-special}
	\begin{tabular}{lccccccccc}
		\hline
		$GRB$ & $T_{\rm 90}$ & $E_{\rm p}$ & $S_{\gamma,6}$ & $F_{\rm p,6}$ & $\rm Association$\\
		& (s) & (keV) & (erg $\rm cm^{-2}$) & (ph $\rm s^{-1}$ $\rm cm^{-2}$) & &  \\
		\hline
		GRB 090618 & $112.39 \pm 1.09$ & $149.04 \pm 3.29$ & $274 \pm 1.51$ & $76.16 \pm 4.75$ & SN \\
		GRB 091127 & $8.7 \pm 0.57$ & $32.73 \pm 4.43$ & $18.3 \pm 0.21$ & $102.97 \pm 2.21$ & SN \\
		GRB 101219B & $51.01 \pm 1.78$ & $82.6 \pm 4.61$ & $2.2 \pm 0.11$ & $3.16 \pm 0.84$ & SN \\
		GRB 111228A & $99.84 \pm 2.11$ & $26.55 \pm 1.37$ & $15.5 \pm 0.38$ & $27.58 \pm 1.74$ & SN \\
		GRB 130215A & $143.75 \pm 13.03$ & $209.95 \pm 42.31$ & $17.6 \pm 0.43$ & $5.18 \pm 1.15$ & SN \\
		GRB 130427A & $138.24 \pm 3.23$ & $824.99 \pm 5.45$ & $1412 \pm 1.82$ & $1259.22 \pm 10.51$ & SN \\
		GRB 130702A & $58.88 \pm 6.19$ & $15.17 \pm 0.28$ & $3.67 \pm 0.06$ & $16.51 \pm 4.69$ & SN \\
		GRB 140606B & $22.78 \pm 2.06$ & $577.63 \pm 102.83$ & $9.57 \pm 0.43$ & $16.26 \pm 1.35$ & SN \\
		GRB 141004A & $2.56 \pm 0.61$ & $181.77 \pm 55.61$ & $1.42 \pm 0.18$ & $17.65 \pm 1.92$ & SN \\
		GRB 150101B & $0.08 \pm 0.93$ & $125.11 \pm 48.57$ & $0.08 \pm 0.02$ & $10.48 \pm 1.35$ & KN \\
		GRB 160821A & $1.09 \pm 0.98$ & $91.97 \pm 27.87$ & $0.17 \pm 0.02$ & $9.16 \pm 1.19$ & KN \\
		GRB 170817A & $2.05 \pm 0.47$ & $215.09 \pm 54.22$ & $0.14 \pm 0.03$ & $3.73 \pm 0.93$ & KN/GW \\
		GRB 171010A & $107.27 \pm 0.81$ & $137.66 \pm 1.42$ & $672 \pm 1.66$ & $137.25 \pm 4.11$ & SN \\
		GRB 190114C & $116.35 \pm 2.56$ & $998.6 \pm 11.9$ & $399 \pm 0.81$ & $344.34 \pm 3.95$ & SN \\
		GRB 200826A & $1.14 \pm 0.13$ & $89.8 \pm 3.7$ & $4.8 \pm 0.1$ & $64.25 \pm 2.05$ & SN \\
		GRB 211211A & $34.31 \pm 0.57$ & $646.8 \pm 7.8$ & $540 \pm 1$ & $476.68 \pm 7.13$ & KN \\
		GRB 230307A & $34.56 \pm 0.57$ & $936 \pm 3$ & $2951 \pm 4$ & $954.33 \pm 16.58$ &  KN \\
		\hline
	\end{tabular}
\end{table*}

Interestingly, we find that the four GRBs (GRB 090618, GRB 130427A, GRB 171010A and GRB 190114C) associated with SNe are clustered on the upper left of the t-SNE and UMAP maps, and three GRBs (GRB 150101B, GRB 160821B and GRB 170817A) associated with KNe are also very close on the t-SNE and UMAP maps.
It is believed that GRB 090618, GRB 130427A and GRB 190114C might originate from binary-driven hypernova (BdHN) \citep{2012A&A...548L...5I,2019ApJ...874...39W}, while GRB 160821B and GRB 170817A originate from the merger of binary neutron stars \citep{2019MNRAS.489.2104T}.
Because relatively few bursts have known counterparts or other additional information, it is difficult to connect these clusters with physical origin.
However, the few bursts known to have a common origin, such as supernovae, kilonovae, are indeed mapped to nearby locations.
These results support that the t-SNE and UMAP maps may indeed aggregate GRBs with the same origin, which provides a way to find or verify GRBs without GW observations as KN candidates in the future.

\subsection{Special GRBs} \label{subsec:special}

\begin{figure*}
	\centering
	\includegraphics[angle=0,scale=0.52]{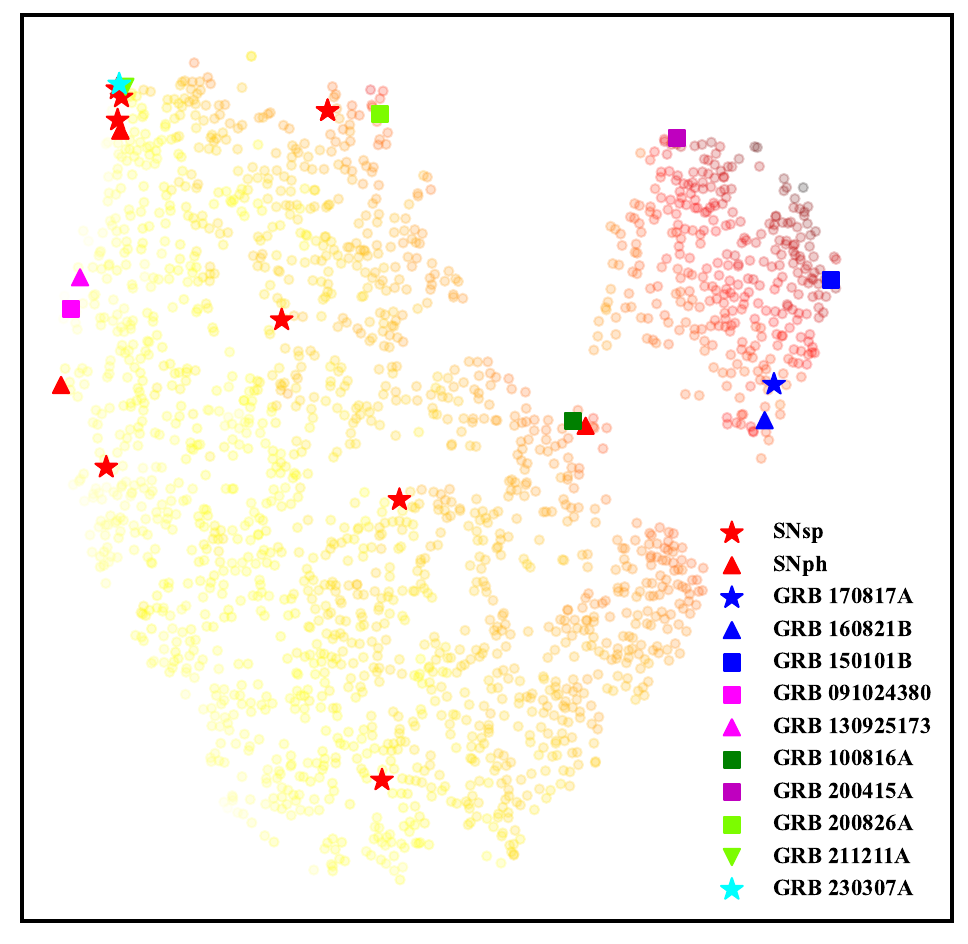}
	\includegraphics[angle=0,scale=0.52]{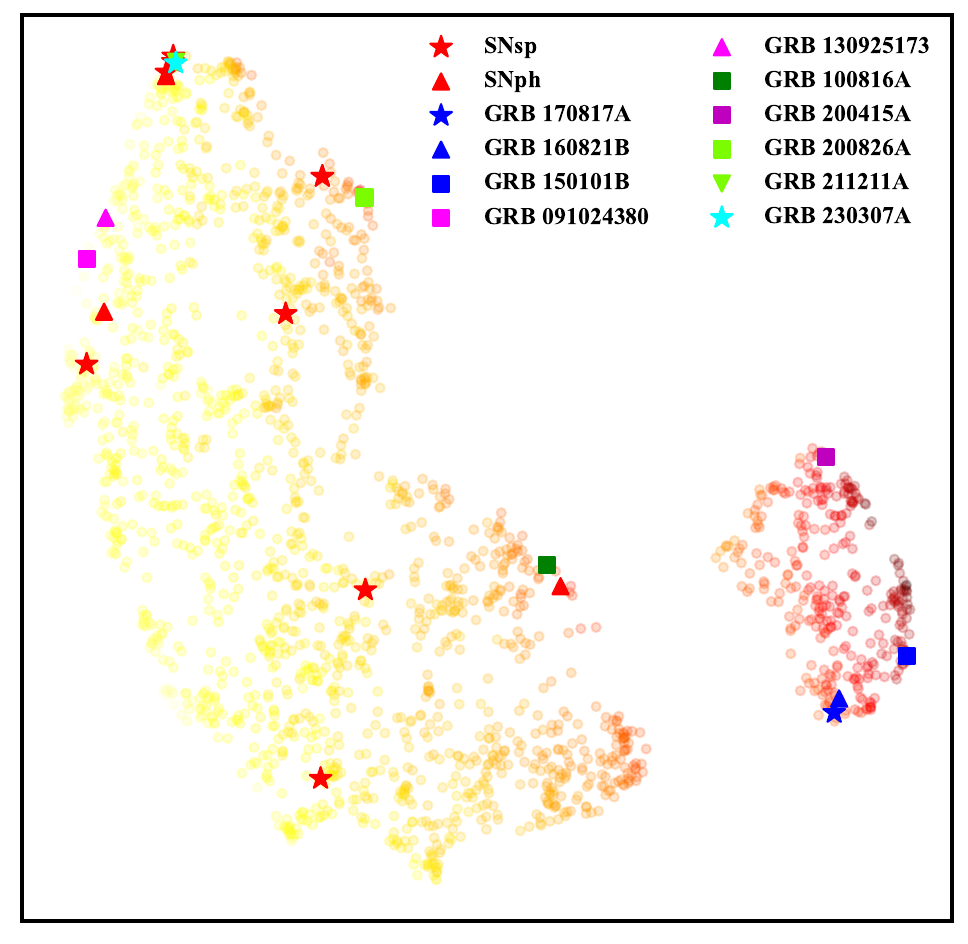}
	\caption{The locations of some special GRBs on the t-SNE (left) and UMAP (right) maps.}
	\label{f-tsne1}
\end{figure*}

GRB 100816A is a controversial GRB, and its duration and spectral lag are consistent with being a SGRB \citep{2010GCN.11113....1N}.
\cite{2011ApJ...739...47F} favored the free-wind medium model for GRB 100816A, in which the progenitor should be a massive star rather than a compact binary.
\cite{2012ApJ...750...88Z} calculated the energy ratio of GRB 100816A and suggested that the GRB 100816A should belong to LGRBs.
In addition, GRB 100816A also complies with the $E_{\rm p,z}$--$E_{\rm iso}$ correlation of LGRBs \citep{2011A&A...531A..20G}.
As shown in Figure \ref{f-tsne1}, GRB 100816A is located at the region of GRBs-II with a GRB associated with SNph near it, which also indicates that GRB 100816A might originate from massive star collapse.

Since no SN association of GRB 100816A was observed with the subsequent observations, the origin of those short LGRBs is still not entirely certain.
However, the observation of GRB 200826A gives decisive evidence that there are some GRBs with $T_{90}<2$ s originating from collapsar.
GRB 200826A is also a unique GRB, which is a typical SGRBs with $T_{90}=1.14$ s, but it has some properties of LGRBs.
The spectral lag of GRB 200826A is 0.157 s, which is at odds with SGRBs but is typical for LGRBs \citep{2021NatAs...5..911Z}; GRB 200826A is fully consistent with the $E_{\rm p,z}$--$E_{\rm iso}$ correlation followed by LGRBs \citep{2020GCN.28301....1S}; \cite{2021MNRAS.503.2966R} favoured that the break in the radio light curve of GRB 200826A originates from the synchrotron self-absorption frequency, which supports that GRB 200826A is a LGRB.
More importantly, GRB 200826A has been found that it is associated with a SN in recent observations, which confirm that GRB 200826A originates from a massive collapsar.
This result further supports the hypothesis that GRB 200826A is actually a LGRB with a duration in the short tail of the distribution \citep{2021NatAs...5..917A,2021NatAs...5..911Z,2022ApJ...932....1R}.
As shown in Figure \ref{f-tsne1}, GRB 200826A is clearly located in GRBs-II.

GRB 091024 is an ultra-long GRB with $T_{90} \approx 1020$ $\rm s$ and triggered Fermi/GBM twice (GRB 091024372 and GRB 091024380) \citep{2009GCN.10070....1B,2011A&A...528A..15G}.
GRB 091024372 is the first emission episode with $T_{90} \approx 94$ s, and GRB 091024380 is the second emission episode with $T_{90} \approx 450$ s. GRB 130925A is also an ultra-long GRB with $T_{90} \approx 4500$ s and triggered Fermi/GBM twice, GRB 130925164, and GRB 130925173, respectively \citep{2009GCN.10070....1B,2011A&A...528A..15G,2014ApJ...781L..19H}.
GRB 130925164 may be the precursor pulse of GRB 130925A with $T_{90} \approx 6$ s, and GRB 130925173 may be the second emission episode with $T_{90} \approx 216$ s \citep{2013GCN.15255....1F,2013GCN.15261....1J}.
As shown in Figure \ref{f-tsne1}, those second emission episodes, GRB 091024380 and GRB 130925173, of these two bursts are very close on the t-SNE map, which implies that they may have a common physical origin.

Recently, a short-hard GRB named GRB 200415A was observed, with a position coincident with the Sculptor Galaxy (NGC 253).
However, its various properties can be explained naturally by the magnetar giant flare, including its location, temporal and spectral features, energy, statistical correlations, and high-energy emissions.
This raises the question of whether GRB 200415A is a classic SGRB or a magnetar giant flare \citep{2020ApJ...903L..32Z,2021Natur.589..207R}.
Note that magnetar giant flares, if occurring in nearby galaxies, would appear as cosmic short-hard GRBs. As shown in Figure \ref{f-tsne1}, GRB 200415A is located in GRBs-I and far away from GRB 170817A, GRB 160821B, and GRB 150101B, which indicate that GRB 200415A may have a different origin.

\subsection{Comparison with the Traditional Short and Long Classification} \label{subsec:different}

For UMAP (t-SNE) result, there are 42 (47) long GRBs-I with $T_{90} > 2$ s, accounting for 12.6\% (13.9\%) of the GRBs-I, and 14 (14) short GRBs-II with $T_{90} < 2$ s, accounting for 0.8\% of the GRBs-II, which are different from the previous traditional short and long classification methods.
GRB 100816A is considered as short GRBs-II due to its particularity, even though its $T_{90}$ is slightly larger than 2 s in our sample.
GRB 170817A is not considered as long GRBs-I , even though its $T_{90}$ is slightly larger than 2 s.

\begin{figure*}
	\centering
	\includegraphics[angle=0,scale=0.52]{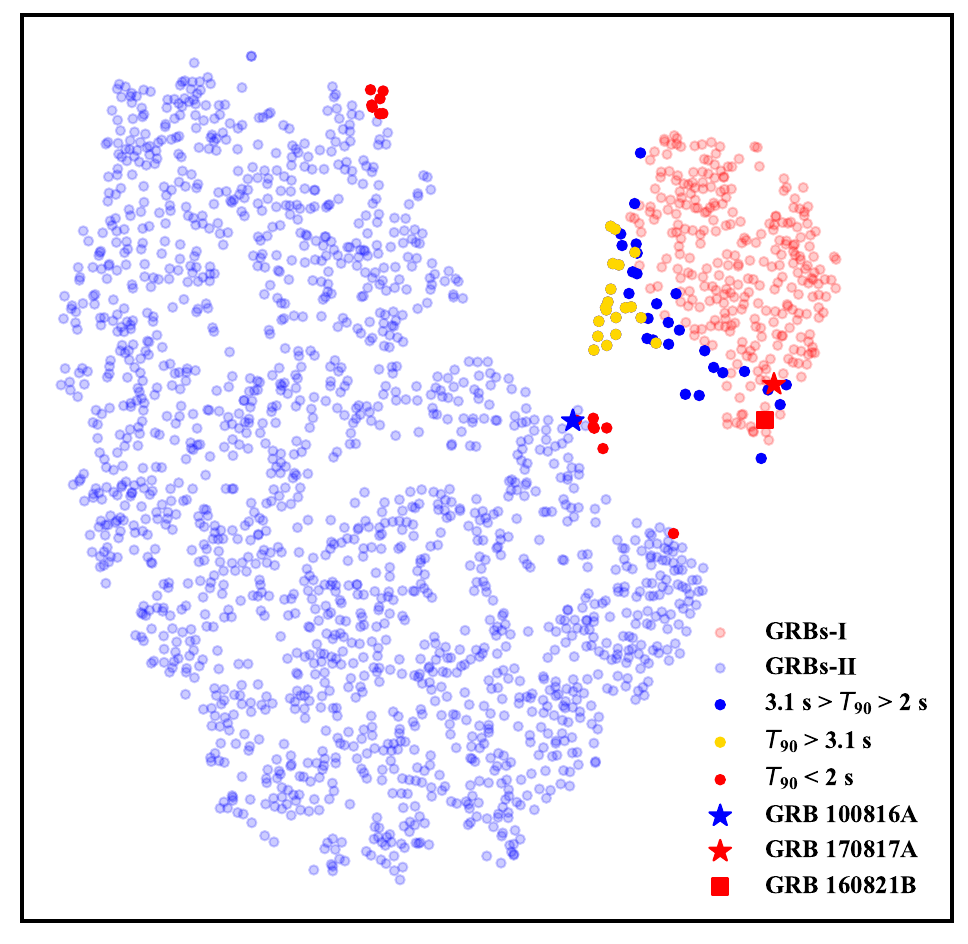}
	\includegraphics[angle=0,scale=0.52]{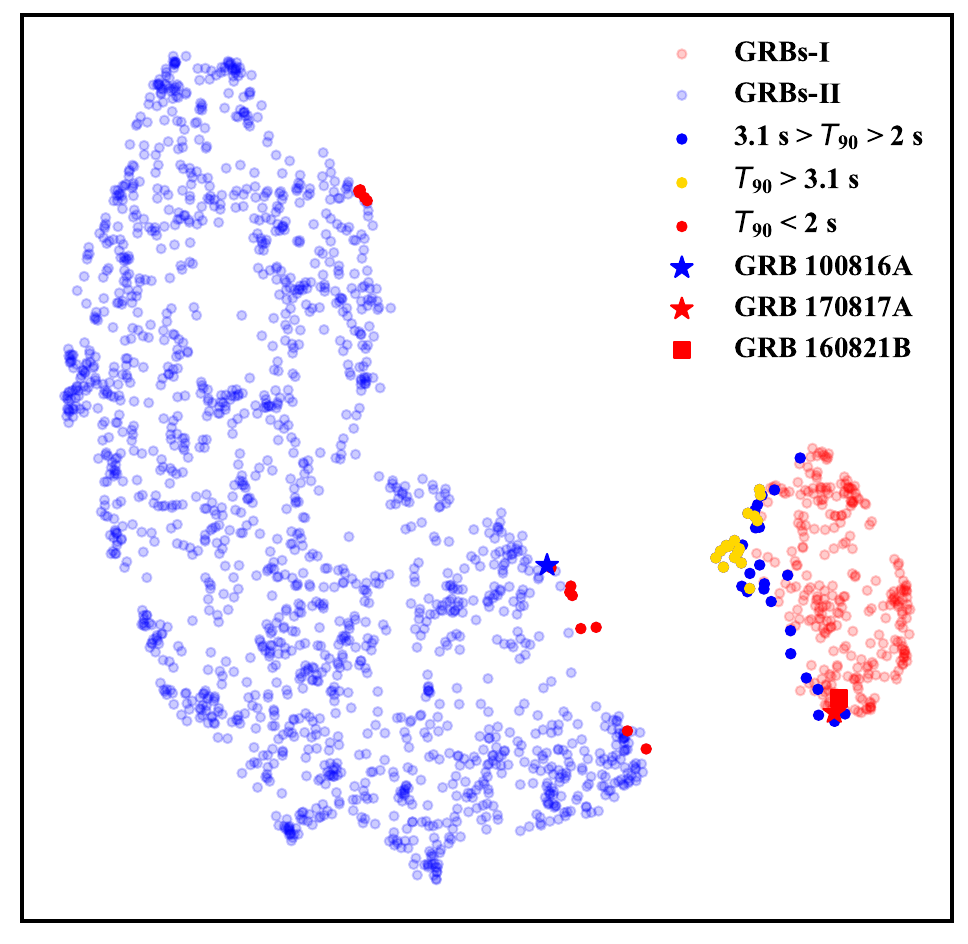}
	\caption{Distributions of long GRBs-I and short GRBs-II on the t-SNE (left) and UMAP (right) maps.}
	\label{f-tsne2}
\end{figure*}

As shown in Figure \ref{f-tsne2}, except for the GRB 180511437 on the t-SNE map and GRB 180511437 and GRB 131128629 on the UMAP map, other short GRBs-II are clustered near GRB 100816A and GRB 200826A.
There are 6 short GRBs-II near GRB 200826A on the t-SNE map, namely GRB 120323A, GRB 140209A, GRB 150819B, GRB 170206A, GRB 171126A and GRB 180703B.
For the UMAP result, GRB 150819B is classified as short GRBs-I and remaining 5 GRBs are also near GRB 200826A.
We find that the spectral lags of GRB 140209A and GRB 180703B \citep{2014GCN.15814....1N,2018GCN.22932....1P} are similar to GRB 200826A and obviously consistent with the traditional LGRBs.
This supports the fact that some ``short" GRBs are produced by the collapse of stars, represented by GRB 100816A and GRB 200826A. Although they may exhibit the properties of ``long" GRBs in some ways, they may be limited by observed conditions, and are misclassified.

We also analyze long GRBs-I and find some evidence that these bursts might be associated with compact star mergers.
GRB 180618A is considered to be a SGRB with extended emission \citep{2018GCN.22794....1H,2018GCN.22822....1S}; GRB 140619B is also considered to be a SGRB and may be from a binary neutron stars merger \citep{2014GCN.16420....1K,2015ApJ...808..190R}; GRB 141222A with the negligible spectral lag is similar to SGRBs \citep{2014GCN.17225....1G}.
Recently, \cite{2024ApJ...963L..12P} analyzed the formation rate of LGRBs and found that low redshift LGRBs could also have compact merger progenitors.

Short GRBs-II and long GRBs-I may be the tails of $T_{90}$ bimodal distribution and the source of the ``intermediate" GRBs.
The t-SNE and UMAP methods can clearly classify them.
For long GRBs-I, some theoretical models such as the merger of NS and massive white dwarfs (WD) \citep{2007MNRAS.374L..34K} have been suggested.

Since there are 4 special GRBs with partial data from GCN, we also perform a stability analysis after removing them. We found that after removing them, the classification results of t-SNE and UMAP changed for only 1 GRB each.
GRB150819440 ($T_{90}$=0.96 s) changed from GRBs-II to GRBs-I in the t-SNE results.
GRB081006604 ($T_{90}$=6.4 s) changed from GRBs-I to GRBs-II in the UMAP results.
However, due to the lack of more information on these two GRBs, they could not be further analyzed.
Therefore, the results suggest that some specific GRBs do affect the classification of individual GRBs, but the effect on the whole sample is negligible.

\section{Conclusions} \label{sec:conclusions}

We use the unsupervised dimensionality reduction algorithms t-SNE and UMAP to classify 2061 Fermi GRBs based upon the four observed quantities, duration, peak energy, fluence and peak flux. The t-SNE ($perplexity=60$) and UMAP ($n\_neighbors=30$ and $min\_dist=0.01$) maps show that these GRBs are clearly divided into two clusters, which support that GRBs should be classified into two types and might indeed originate from two different progenitors. We label the two clusters as GRBs-I and GRBs-II, and find that all GRBs associated with supernovae are classified as GRBs-II, especially for GRB 200826A, a peculiar short GRB that has been confirmed from a collapsar. Furthermore, except for GRB 211211A and GRB 230307A, all GRBs associated with kilonovae belong to GRBs-I. Three GRBs (GRB 090618, GRB 130427A and GRB 190114C) originated from binary-driven hypernova are clustered on the upper left of the t-SNE and UMAP maps, and two GRBs (GRB 160821B and GRB 170817A) originated from the merger of binary neutron stars are also very close on the t-SNE and UMAP maps. These results indicate that the t-SNE and UMAP methods may indeed aggregate GRBs with the same origin, and GRBs-I are associated with compact mergers while GRBs-II are associated with collapsars.

Contrary to the traditional short and long classification, there is no absolute boundary of duration between GRBs-I and GRBs-II does not have an absolute boundary.
$T_{90}$ of GRBs-I can reach a maximum of 8 s, and $T_{90}$ of GRBs-II can reach a minimum of 0.4 s. Furthermore, we find that more than 10\% of GRBs-I duration is greater than 2 seconds and about 1\% of GRBs-II duration is less than 2 seconds. This indicate that part of short bursts have different origin, perhaps from collapsing stars, which has been confirmed by GRB 200826A. While some long bursts have also different origins than most, probably from compact star mergers, such as GRB 211211A and GRB 230307A.

\section{acknowledgments}
We thank the anonymous referee, Hou-Jun L{\"u} and Lang Shao for insightful comments/suggestions. We also acknowledge the use of public data from the Fermi catalogue. This work was supported in part by the Guangxi Natural Science Foundation (No. 2022GXNSFDA035083), and by the National Natural Science Foundation of China (No. 11763003).

\section*{Data Availability}
The Fermi GRB data underlying this article are publicly available at the Fermi Catalog (https://heasarc.gsfc.nasa.gov/W3Browse/fermi/fermigbrst.html).

\bsp	
\label{lastpage}
\end{document}